\documentclass[
preprint,
superscriptaddress,
aps,
prl,
]{revtex4-1}

\usepackage{graphicx}
\usepackage{dcolumn}
\usepackage{bm}
\usepackage[utf8]{inputenc}
\usepackage{amsmath, amsthm, amssymb, amsfonts}
\usepackage{accents}
\usepackage[usenames,dvipsnames]{xcolor}
\usepackage[caption=false]{subfig}
\usepackage{tikz}
\usepackage{array}
\usepackage{dsfont}
\usepackage{mathtools}
\usepackage{soul,xcolor}
\usepackage[capitalise]{cleveref}

\usepackage{color}
\usepackage{latexsym,amstext}
\usepackage{url}

\pgfdeclarelayer{bg}    
\pgfsetlayers{bg,main}  

\usetikzlibrary{shapes.geometric}
\usetikzlibrary{shapes.misc}
\usetikzlibrary{arrows.meta}

\newcolumntype{M}[1]{>{\centering\arraybackslash}m{#1}}

\setstcolor{blue}

\begin{document}

\newcommand{\mytitle}{\Large Antagonistic Phenomena in Network Dynamics}
\title{\mytitle}

\author{Adilson E. Motter}
\affiliation{$\mbox{\footnotesize Department of Physics and Astronomy, Northwestern University, Evanston, Illinois 60208, USA.}$}
\affiliation{$\mbox{\footnotesize Northwestern Institute on Complex Systems, Northwestern University, Evanston, Illinois 60208, USA.}$}
\author{Marc Timme}
\affiliation{$\mbox{\footnotesize Chair of Network Dynamics, Institute for Theoretical Physics and Center for Advancing Electronics (cfaed),}$ 
$\mbox{\footnotesize Technical University of Dresden, 01062 Dresden, Germany.}$}
\affiliation{$\mbox{\footnotesize Network Dynamics, Max Planck Institute for Dynamics and Self-Organization, 37077 G\"ottingen,  Germany.}$}

\begin{abstract}
Recent research on the network modeling of complex systems has led to a convenient representation of numerous natural, social, and engineered systems that are now recognized as networks of interacting parts. Such systems can exhibit a wealth of phenomena that not only cannot be anticipated from merely examining their parts, as per the textbook definition of complexity, but also challenge intuition even when considered in the context of what is now known in network science.  Here we review the recent literature on two major classes of such phenomena that have far-reaching implications: (i) antagonistic responses to changes of states or parameters and 
(ii)~coexistence of seemingly incongruous  behaviors or properties---both deriving from the collective and inherently decentralized nature of the dynamics. They include effects as diverse as negative compressibility in engineered materials, rescue interactions in biological networks, negative resistance in fluid networks, and the Braess paradox occurring across transport and supply networks. They also include remote synchronization, chimera states and the converse of symmetry breaking in brain, 
power-grid and oscillator networks as well as remote control in biological and bio-inspired systems.  By offering a unified view of these various scenarios, we suggest that they are representative of a yet broader class of unprecedented network phenomena that ought to be revealed and explained by future research.
\end{abstract}

\hfill{\footnotesize  Annu. Rev. Condens. Matter Phys. 9, 463 (2018)}

\maketitle
\section{1.~INTRODUCTION}

Many systems in nature and society can be conceptualized as a collection of parts coupled through a web of interactions and suitably modeled as a network. A network can be represented as 
 {\color{black} a} collection of nodes---e.g., particles, genes, or individuals---connected by links reflecting the interactions between them. Two main lines of research have contributed to the study of such systems:  reductionist studies, focused on separating the parts to determine their individual properties; and complex systems research, focused on the whole-system behavior. The emerging field of network science 
  {\color{black} owes}
  its success to the recognition that collective behavior is largely determined by the network of interactions between the parts. Substantial attention in this field has been dedicated to the study of network structure alone and to efforts to attribute as much as possible of  the observed collective behavior to the properties of this structure. 
Attempts to infer the collective behavior 
solely from the properties of the parts
 {\color{black} are}
prone to failure, and this has long been appreciated in condensed matter physics and other fields, which nevertheless does not make those properties unimportant.  Here we consider phenomena that 
 {\color{black} depend}
on {\it both} the network structure {\it and} the properties of the parts, and 
 {\color{black} are}
thus determined by the interplay between the network structure and dynamics. 

The first part of this article is focused on scenarios in which the {\it removal}  of resources from a network---e.g., through the removal of nodes and/or links---can in fact {\it improve} network function or performance. 
The notion that ``less can be more'' has been long appreciated in connection with minimalism in architecture and arts, the paradox of choice in psychology, diminishing returns in economics, calls to stop the seemingly endless gadget feature explosion, 
and even the rise of microblogs such as Twitter. What is not widely appreciated (albeit common, as we argue) is that a similar notion could govern complex networks, which in many cases have evolved to have more (not less) nodes and links. 

\begin{figure}[h]  
\includegraphics[width=3.4in]{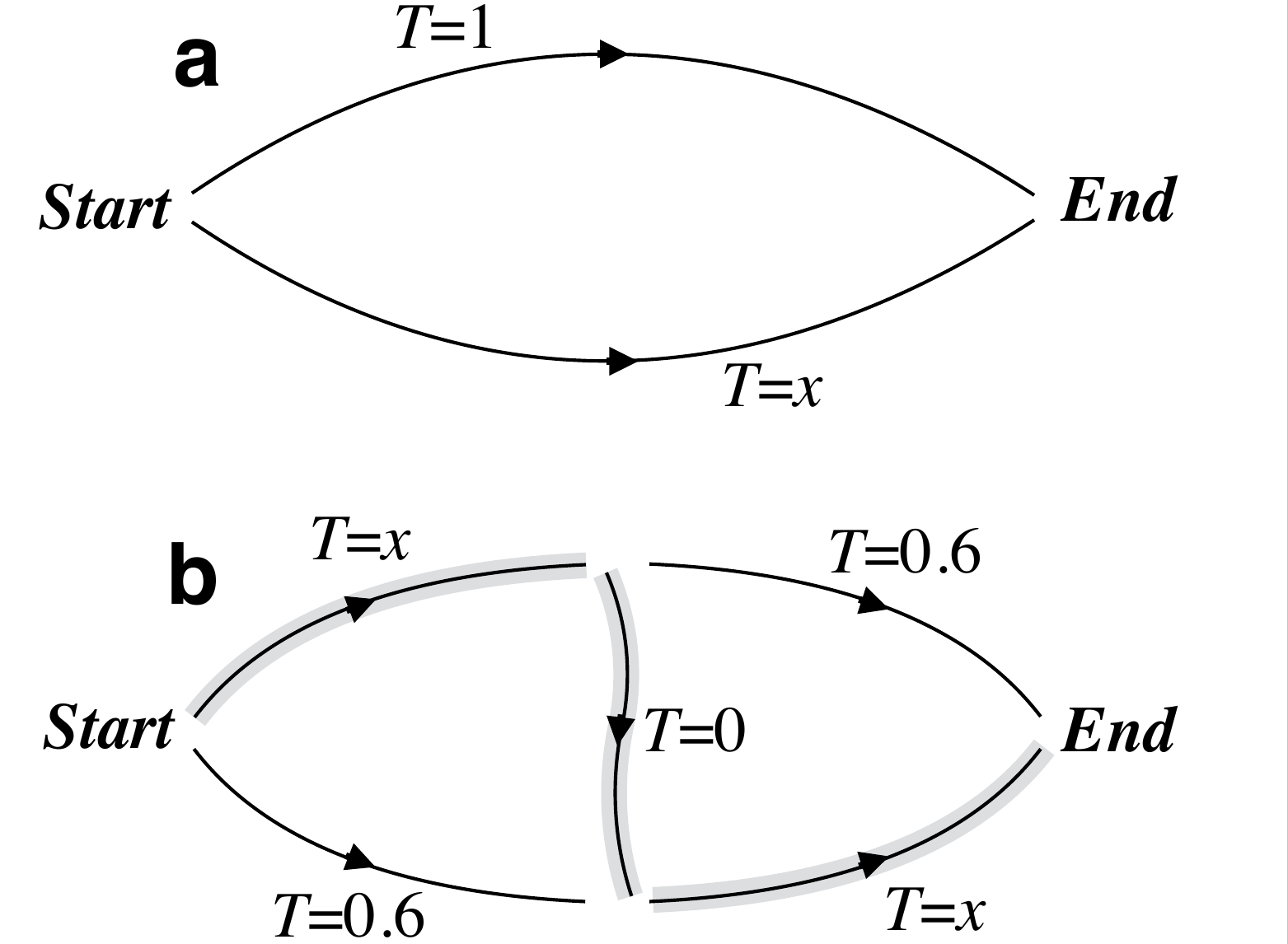}
\caption{Traffic  paradoxes, where $T=T(x)$ is the travel time as a function of the fraction of all users $x$ taking the particular path.
(a) Pigou-Knight-Downs paradox, 
 where, in equilibrium,
 all users take the lower path even though the 
 globally optimal solution would split traffic evenly.
In this example, the path with fixed travel time (top path) is often interpreted as public transportation, whereas the path with traffic-dependent  travel  time (bottom path) can be interpreted as car roads. 
(b) Braess paradox, where all users take the marked path through the $T=0$ shortcut. In this case, global optimization of travel time requires the traffic to be equally 
distributed between the top and bottom paths, which is achieved by removing the shortcut road.
The non-cooperative routing causes the average travel time  to increase from  $0.75$ to $1$ in (a) and from $1.1$ to $2$ in (b).}
\label{fig1}
\end{figure}

The key network property underlying such phenomena is that the equilibrium state spontaneously reached by a decentralized network is not necessarily the global 
{\color{black} optimum}
of the system. Thus, even though the removal of resources 
constrains the solution
space, which cannot improve (and generally 
 {\color{black} worsens)}
the optimum of the objective function, it can counterintuitively do so while bringing the equilibrium state closer to the optimum. 
In the economics literature this mechanism has been known for at least a century \cite{pigou1932economics}, and it now forms the basis of the Pigou-Knight-Downs paradox, which describes scenarios  in which investment in roads does not improve door-to-door equilibrium speed because it incentivizes people to shift from public transportation to driving (see \textbf{Figure \ref{fig1}a}).
The mechanism was 
rediscovered in different contexts over the years (likely independently),
and by the 1950's variants of it had been reported in the mainstream transportation literature {\color{black} \cite{wardrop1952road,beckmann1956studies}.} In transportation, the best-known formulation comes from the work of Braess, who in 1968 described what is now known as the Braess paradox 
\cite{Braess1969, braess2005paradox}.  
{\color{black}
The paradox arises when}
 the addition of 
 an intermediate road to a traffic network---which effectively increases
 its   
 capacity---has the consequence of increasing rather than decreasing
 the average travel time between origin and destination even if the total number of cars remains the same  (see \textbf{Figure \ref{fig1}b}). Related concepts have been explored in computer science and operations research, where the difference between the equilibrium and the optimum of the objective function (i.e., the travel time in the example just given) is often called the price of anarchy \cite{roughgarden2005selfish}.

This class of problems also admits a natural formulation in game theory, where 
they can be formally related to social dilemmas, namely situations in which an agent profits from being selfish unless everyone chooses to be selfish, in which case everyone loses.
Indeed, in the example of \textbf{Figure~\ref{fig1}b},
it is the selfish routing \cite{roughgarden2005selfish} of the drivers, who seek to optimize their own travel time with no regard to the travel time of the others, that causes the shift of the system to a less desirable equilibrium characterized by a longer travel time. Once the shift has occurred, no faster route is available to a driver  no matter what individual choices 
the driver
{\color{black} may}  
make. In the literature of non-cooperative games \cite{nash1950equilibrium}, this scenario is known as a Nash equilibrium and is described as a stable state in which no agent can gain from a unilateral change of strategy. 
In this language, the Braess paradox emerges from the fact that the Nash equilibrium is not necessarily optimal and thus  
a capacity increase can further lower its fitness.  As discussed below, the recent network literature shows that analogous behaviors may be generic in many physical and biological networks, where they give rise to a wealth of seemingly disparate phenomena.

The second part of the article is centered on network phenomena that invoke coexistence of seemingly incompatible properties or 
 {\color{black} behaviors.}
One may figuratively argue that things that occur and stay together must ultimately fit together.
In networks, however, the verification of this principle is far 
from obvious. 
For example, the time evolution of two isolated chaotic systems diverges even if they are identical---owing to their inherent sensitive dependence on initial conditions---but when weakly coupled, they can  synchronize stably to the exact same {\color{black} trajectory,}
 which may even be a solution of their isolated dynamics \cite{pecora1990synchronization}. Using synchronization as a model process 
  {\color{black} of}
  behavioral uniformity that can emerge from interactions, below we discuss a selection of phenomena that seem implausible in the absence of detailed analysis.

Before proceeding we note that although it is almost impossible to talk about collective behavior without thinking of the 
  {\color{black} notion}
popularized by Anderson that ``more is different'' \cite{anderson1972more}, the ideas covered here are closer in content to 
  {\color{black} Watts'}
notion that the common sense to which our intuition has been trained is not a valid scientific tool in the study of network systems \cite{watts2011everything}. He argues for the need of a form of ``uncommon sense,'' which a rigorous wholistic  network-based description that integrates structure and dynamics would conceivably help provide. A key difference from Watt's description is that 
his examples are mostly in the context of social sciences, where 
a failure of intuition can be partly attributed to unknowns, while 
the phenomena we describe here manifest themselves in a form that is counterintuitive even when the mathematical description is assumed to be known, exact, complete, and deterministic.

\section{2.~ANTAGONISTIC DYNAMICS}

\begin{figure}[h]
\includegraphics[width=1.7in]{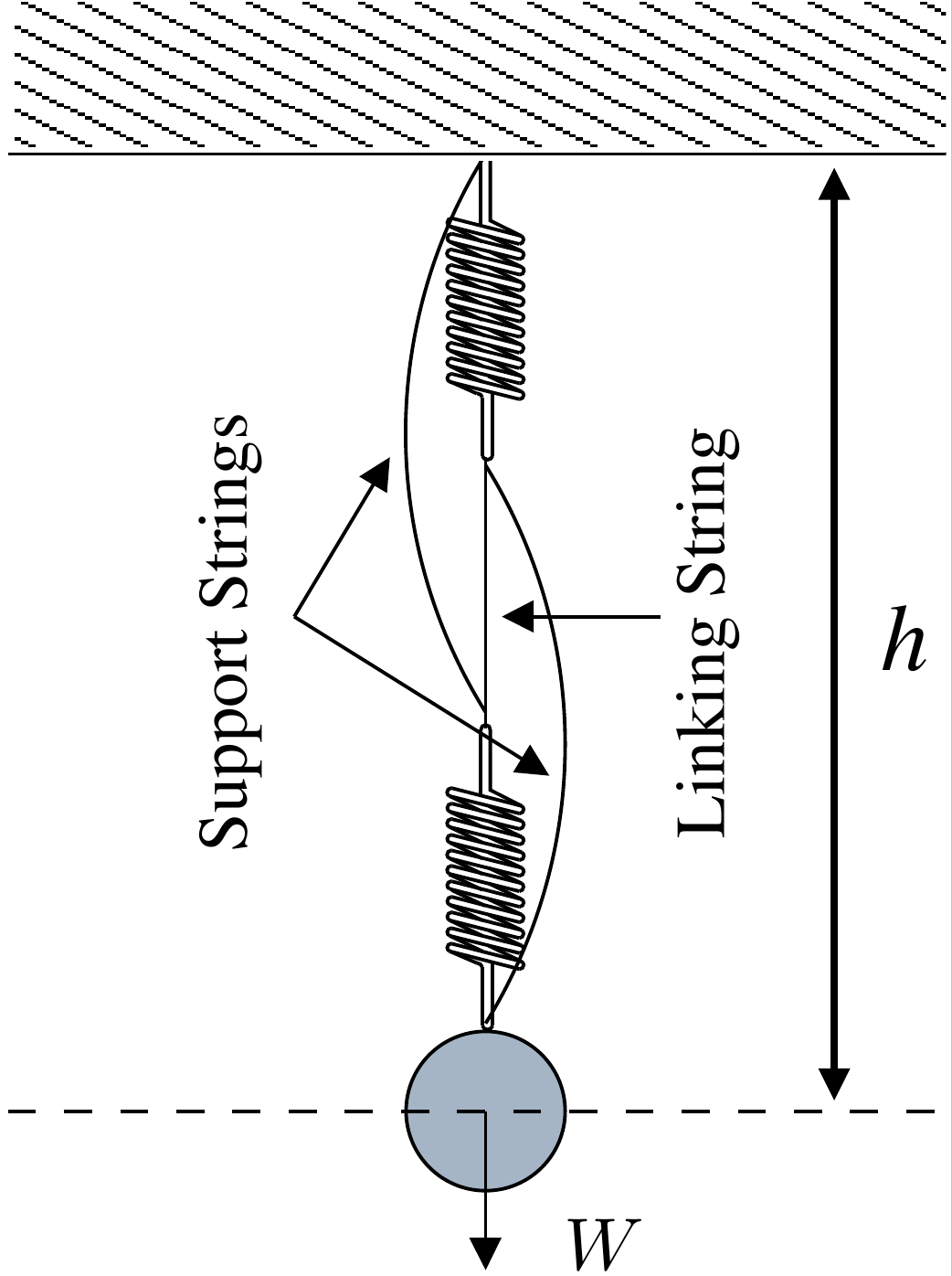}
\caption{Mechanical analog of the Braess paradox, where the removal of the linking string can cause the equilibrium position 
of the weight $W$ to rise. Figure based on Reference \cite{cohen1991paradoxical}.
}
\label{fig2}
\end{figure}

Consider a weight $W$ supported by a system of two
{\color{black} identical}
 springs connected by a linking string, and assume the setup includes two {\color{black} identical} slack support strings, as shown in  \textbf{Figure~\ref{fig2}}.
If the linking string is cut, the support strings become taut and, contrary to the common-sense expectation, the weight can rise.  This occurs because the removal of the linking string causes the springs to go from a series configuration to a parallel configuration, prompting them to contract. Indeed, initially each spring holds the entire weight $W$, leading to an equilibrium height 
{\color{black} 
$h= 2l_0+l_\ell+2 \frac{W}{k}
= l_0 + l_s-\delta+\frac{W}{k}$,} where $k$ is the spring constant,  $l_0$ is the length of {\color{black}  the individual} unstretched springs,   
{\color{black} $l_\ell$ is the length of the linking string,
$l_s$ is the length of the individual} support strings, 
and $\delta$ is the amount of slack the {\color{black} support} strings have.
{\color{black} After}
the removal of the linking string, each spring holds only half of the weight $W$, which leads to an equilibrium height 
{\color{black} $h=l_0 + l_s+\frac{W}{2k}$.} It follows that the weight rises by $\Delta{}h=\frac{W}{2k}-\delta$ when the linking string is removed, which is positive if $\delta$ is chosen to be smaller than $\frac{W}{2k}$. 

This behavior has long been known and was popularized in Reference \cite{cohen1991paradoxical}. It can be regarded as a mechanical network effect that is formally equivalent to the Braess paradox  discussed above. This mechanical analog illustrates both that the conditions underlying the paradox can occur in disparate network systems and that the resulting effect can lead to a rich variety of otherwise unrelated network phenomena. We now turn to network problems of significant current interest that expand on these points.

\subsection{2.1.~Synthetic Rescues in Biological Networks}

\begin{figure}[t]
\includegraphics[width=4.6in]{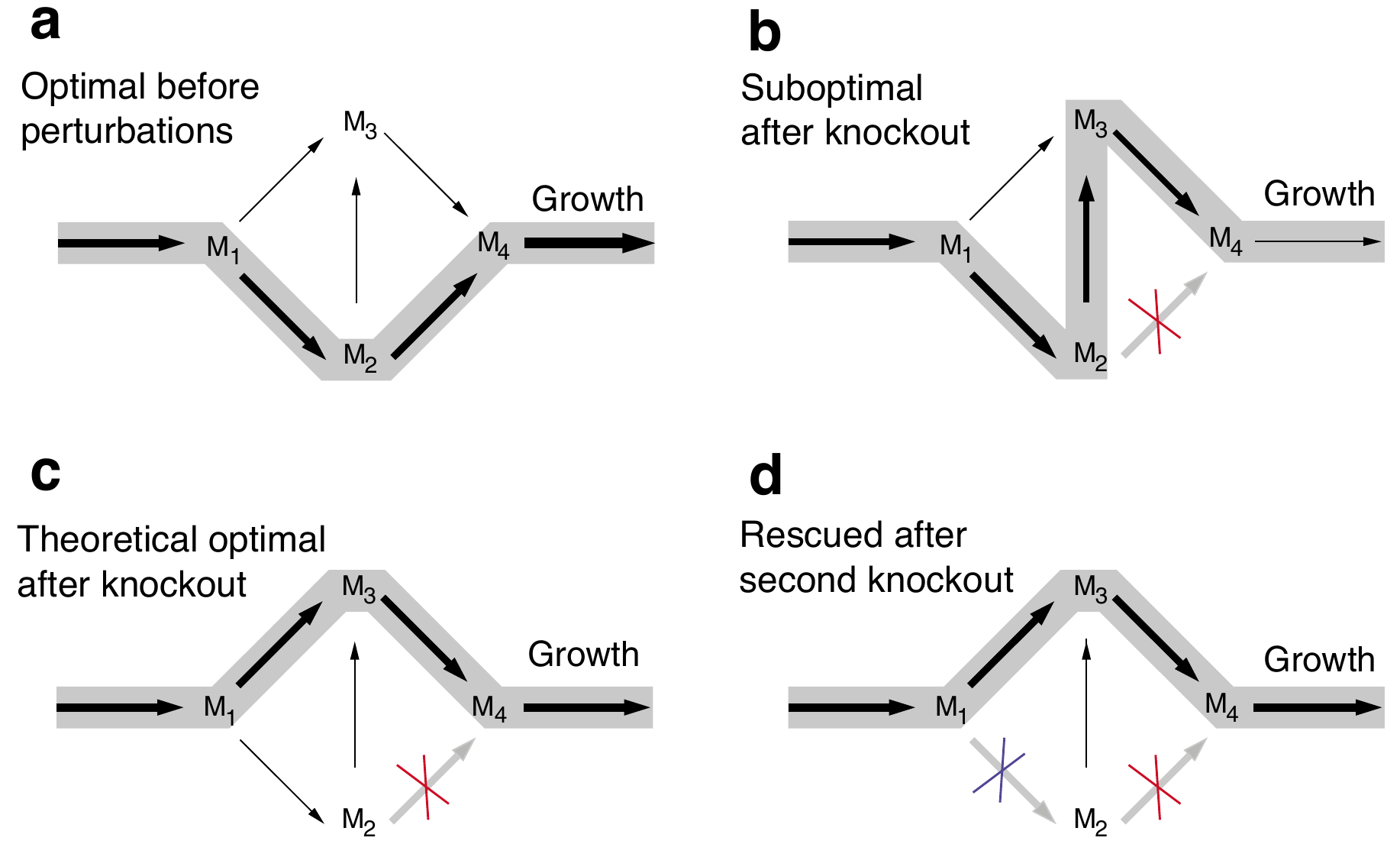}
\caption{Synthetic rescue in a toy metabolic network with four  compounds, $M_1$, $\cdots$, $M_4$, where the width of the arrows 
indicate the strength of the corresponding reaction fluxes  and the shades mark the highest flux pathways.
The panels show the equilibrium metabolic state (a) prior to any perturbation, (b) after 
the primary knockout and
(d) after the rescue knockout, as well as (c)  the optimal growth state after the primary knockout.
Figure adapted from Reference \cite{motter2008predicting}.
}
\label{fig3}
\end{figure}

In living cells, the loss of biological function caused by the inactivation of a gene can sometimes be compensated by the inactivation of additional genes. 
This phenomenon, which has been confirmed experimentally,  was first predicted  in  Reference \cite{motter2008predicting}  in the context of metabolic networks and can be seen as a biological analog of the Braess paradox. For concreteness, consider a single-cell organism for which the biological function of interest is growth (i.e., reproduction) rate, and assume that the cells are fully adapted to their environment, meaning that they maximize growth rate under the given conditions. The knockout of otherwise active metabolic genes leads to the inactivation of the associated metabolic reactions to which the proteins coded by those genes serve as catalyzers. Following such a perturbation, the cells are generally no longer in an optimal growth state. An optimal  state may nevertheless be approached when certain additional genes (hence metabolic reactions) are knocked out, which gives rise to the predicted synthetic rescue, as illustrated in  \textbf{Figure~\ref{fig3}}.

As a minimal description of the phenomenon,  we can use flux balance analysis \cite{edwards2000escherichia} to model the optimal growth state of the cell as a linear programming problem.
In this problem, one seeks to maximize the rate of a putative reaction $v_{bio}$  that models the overall biomass production (and hence growth rate) under the constraints imposed by the stoichiometry of the metabolic network, nutrient availability, thermodynamics, and any imposed reaction inactivation: 
\begin{equation}
\max_{\{v_j\}}~  v_{bio}  \mbox{ ~subject to~   $\sum_{j}$} S_{ij} v_{j} = 0  \mbox{ ~$\forall i$} \mbox{ ~and~ } v_j^{\min} \le v_j \le v_j^{\max} \mbox{ ~$\forall j$},
\label{eq_1}
\end{equation} 
where $S_{ij}$ are the entries of the stoichiometric matrix and $v_j$ are the reaction fluxes.
The suboptimal response to the knockout of a gene can be modeled in its simplest form using the 
{\color{black} ``minimization of metabolic adjustment'' hypothesis}
\cite{segre2002analysis}, which can be implemented as a  
 {\color{black} quadratic} 
programming problem.
{\color{black}
The model}
 identifies the available state ${\boldsymbol  v'}= (v'_j)$ in the space of metabolic fluxes  that, under the additional constraint imposed by the gene inactivation, is closest to the pre-knockout state  ${\boldsymbol  v}$:  $  \min_{\{v'_j\}} || {\boldsymbol v'} -{\boldsymbol  v}||^2  $ subject to constraints in the form of those in Equation \ref{eq_1}. 
 Thus, 
this model
effectively predicts that metabolic fluxes are rerouted mostly locally, whereas the new optimal state could require a more global flux rearrangement. The rescue state can then be predicted by applying the same quadratic optimization to the combined perturbation of the primary and rescue gene knockouts. From a biological standpoint, the second knockout effectively precipitates adaptation of the perturbed network that could in principle be eventually achieved by long-term adaptive evolution~\cite{cornelius2011dispensability}. 

An interesting potential application of this phenomenon is to the development of pairs of antibiotic drugs that can select against resistant cells. Each gene knockout of a synthetic rescue pair has the potential to inhibit growth when implemented in isolation, but one of them suppresses the impact of the other when applied concurrently. Thus, they serve as targets for antibiotic drugs that would interact antagonistically \cite{motter2010improved}, and previous research has shown that the combination of two antagonistically interacting antibiotics  will select against cells that have developed resistance to the suppressor~\cite{chait2007antibiotic}.

\subsection{2.2.~Metamaterials with Negative Compressibility}

\begin{figure}[h]
\includegraphics[width=5.3in]{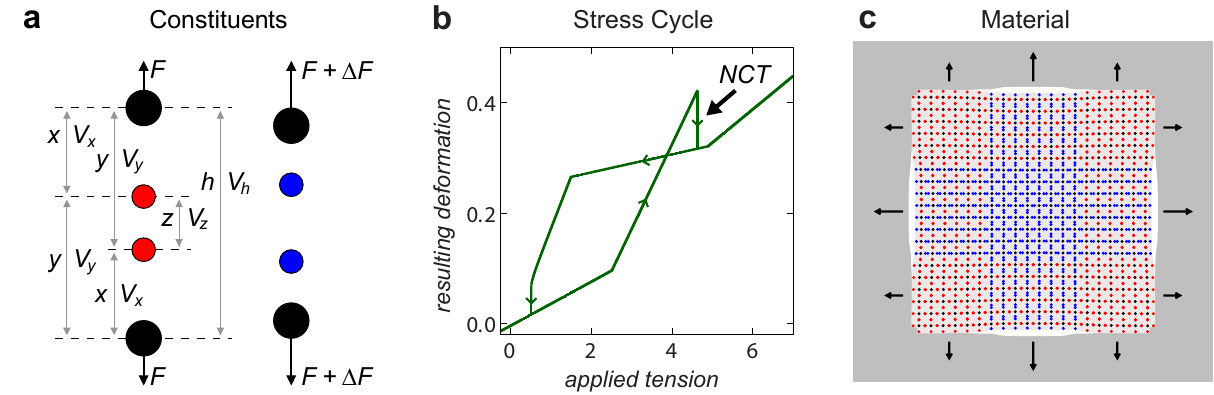}
\caption{Metamaterial exhibiting  negative compressibility transitions.
(a) Constituents before (left) and after (right) contracting in response to increased tension. 
(b) Stress-strain projection of the hysteresis loop {\color{black} of the constituents} as the applied tension is varied. 
{\color{black} The}
negative compressibility transition, 
{\color{black} corresponding to a change from the configuration  on the left to the configuration on the right in panel (a),} is indicated 
on the figure by its initials (NCT).
(c) Material formed by a square lattice of constituents, where the color code is the same as in (a) 
and the white background indicates the extent of the effect.
Figure adapted from  Reference \cite{nicolaou2012mechanical}.
}
\label{fig4}
\end{figure}

While ordinary materials expand when tensioned, it has been shown that a material can be designed to undergo a negative compressibility transition (i.e., a transition to a contracted phase) in response to increasing tension \cite{nicolaou2012mechanical}. As in the case of other  metamaterials, such a  material is engineered to gain its unusual property from its structure rather than  composition. 
{\color{black} In other words,}
this is a property of the underlying mechanical network, 
which can in fact be seen as a nontrivial generalization of the spring-string system discussed above.  To understand this generalization, we can imagine the constituents of the material as consisting of four particles that interact with each other through attractive forces, which are nevertheless nonlinear, allowing the system to be bistable.
{\color{black}  When two stable states coexist, one generally corresponds to an expanded configuration while the other corresponds to a contracted one.}
Ignoring dissipation for simplicity, each such constituent is characterized by a potential of the form 
\begin{equation}
V(x,y,h,F)=V_x(x)+V_y(y)+V_z(y-x)+V_x(h-y)+V_y(h-x)+V_h(h)-Fh,
\label{eq2}
\end{equation}
where $h$ is the total length and $F$ is the applied tensional force (see \textbf{Figure \ref{fig4}a}). As $F$ is increased, the stable state corresponding to larger $h$  first becomes marginally stable and then disappears. If the system was initially in this extended state, it will then transition to the other stable state, which corresponds to a smaller $h$. 

Mathematically, this transition is determined by a bifurcation analogous to one observed for the potential $U(\xi, { f})=-\xi^3/3+{ f}\xi$, where ${ f}$ is the tunable parameter. This potential has a stable equilibrium point at $\xi^*=-\sqrt{ f}$ and an unstable one at $\xi^*=\sqrt{ f}$ for ${ f}>0$, has a single (degenerate) equilibrium point at $\xi^*=0$ for ${ f}=0$, and has no equilibrium point for ${ f}<0$. Thus, as ${ f}$ is varied from positive to negative, the stable equilibrium point vanishes and the system responds discontinuously. Physically, the disappearance of the occupied stable state is analogous to the cutting of the linking string in  \textbf{Figure \ref{fig2}}: it causes the inner particles in  \textbf{Figure \ref{fig4}a} to move closer
{\color{black} to}
 (and hence attract more strongly) 
 the external particles, which is similar to the transition from a parallel to a series configuration in the spring-string system. A key difference here is that the process can be cycled by varying the tensional force, as shown in the hysteresis diagram of \textbf{Figure \ref{fig4}b}.

The material itself can be formed by aggregating such constituents, as shown in \textbf{Figure \ref{fig4}c}  for a square lattice network. In the thermodynamic limit, the bifurcations undergone by the constituents give rise to a transition between the corresponding extended and contracted phases, which can be rigorously characterized at finite temperature using tools from statistical physics \cite{nicolaou2013longitudinal}. Such materials can find  applications in the design of micro-mechanical controls and protective mechanical devices, but they also suggest a more general principle to design metamaterials with inverted responses that can in theory be applied not only to stress-and-strain  but also to any pair of thermodynamically conjugated variables.

\subsection{2.3.~Synchronization Improvement  by Interaction Pruning}

\begin{figure}[h]
\includegraphics[width=5.5in]{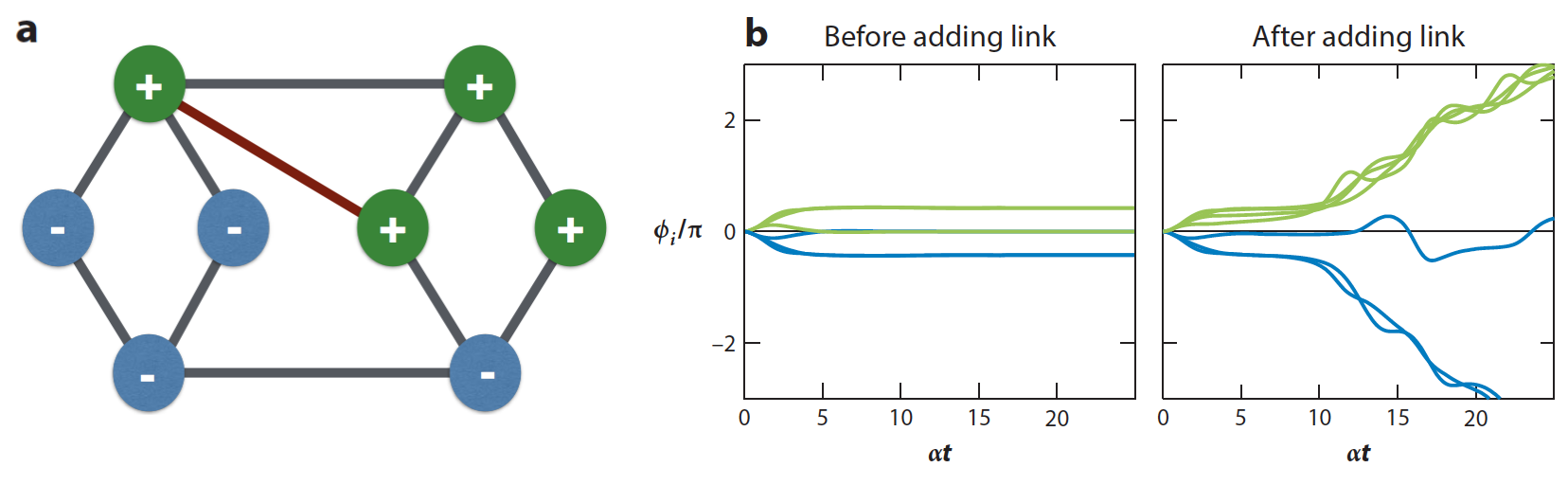}
\caption{Desychronization induced by line addition in a toy power grid governed by  Equation \ref{eq3}.
(a) Example system consisting of four generators (green) and  four motor nodes (blue). A  stable synchronous state 
exists for the network formed by the black lines, but synchronization is lost upon addition
of the red link. 
After this addition, the  sum of the angle differences  
 given by the relation $\phi_i-\phi_j = \Delta_{ij}$ 
violates the condition of being a multiple of  $2\pi$  along some closed loops,  
meaning that the synchronization ansatz used to derive the phases no longer holds.
(b) Time evolution of the phases of the various nodes (color-coded as in  panel (a)) converging to a synchronous state in the absence of the red link (left) and progressing to a de-synchronous state after the link addition (right).
Figure adapted from Reference~\cite{witthaut2012braess}.
}
\label{fig5}
\end{figure}

{\color{black}
In the synchronization of coupled oscillators, a parallel with the Braess paradox has been established in which the addition or strengthening of connections between oscillators has the adverse effect of removing a previously existing  
synchronous state.  This possibility has attracted special attention in connection with power-grid networks, where the addition of new line capacity for power transmission could eliminate a phase-locked operating state among power generators and motors in an AC network \cite{witthaut2012braess,coletta2016linear}. A minimal model to illustrate this effect is 
 \begin{equation}
\ddot{\phi}_i=P_i-\alpha \, \dot{\phi}_i-\sum_j K_{ij} \sin (\phi_i-\phi_j), ~~ i = 1, \cdots, n,
\label{eq3}
\end{equation} 
where both generator nodes ($P_i>0$) and 
motor nodes ($P_i<0$) 
are represented as damped second-order  phase oscillators and $K_{ij}$ represents the network structure as well as the capacity of the transmission lines \cite{filatrella2008analysis,rohden2012self}.
Of special interest are the frequency-synchronized states satisfying $\dot{\phi}_1=\dot{\phi}_2=\dots =\dot{\phi}_n$ at all times and thus $\Delta_{ij}=\phi_i-\phi_j=\textsf{const}$ for all $i$ and $j$. 

This synchronization condition leads to equations of the form $\sum_j K_{ij} \sin \Delta_{ij}=P_i$, where $\Delta_{ij}$ is to be determined for given  $K_{ij}$ and $P_j$.  
When  such a solution for  $\Delta_{ij}$ exists,  the resulting power flows through the lines are $K_{ij} \sin \Delta_{ij}$ and they automatically respect the line capacities.  
A new solution for $\Delta_{ij}$ 
may exist if line capacities are increased (or, in particular, if
new lines are added), 
but therein lies the rub: for the actual {\it state} to exist, not only a solution for each $\Delta_{ij}$ has to exist but also each phase angle $\phi_i$ has to be uniquely defined,  and the latter is not guaranteed when line capacities 
 are increased 
 even if the $\Delta_{ij}$ solution continues to exist.
  That is, the set of equations $\phi_i-\phi_j = \Delta_{ij}$  (which must be simultaneously satisfied for every pair of nodes $i$ and $j$ connected by a  transmission 
 line with non-zero capacity $K_{ij}$) is no longer guaranteed to have a solution. 
 Indeed, as demonstrated in References \cite{witthaut2012braess,witthaut2013nonlocal}, 
 a capacity increase will frequently induce  conflicts between the 
 phases in this set of equations, which are reminiscent of the phenomenon of geometric frustration as the conflicts necessarily occur along loops in the network.
 (In the familiar case of geometric frustration in spin systems, not all pairwise interaction energies can be minimized simultaneously
 precisely due to geometric constraints similar to the ones considered here.) 
 This effect is illustrated in \textbf{Figure \ref{fig5}} for a simple network of four identical generators ($P_i=1$), four identical 
 motors ($P_i=-1$), and lines with identical capacity ($K_{ij}=1.03$ for any line with nonzero capacity).
}

In networks of diffusively coupled oscillators, which describe the dynamics of power generators in the vicinity of their synchronization manifold \cite{motter2013spontaneous}, an analogous effect can be observed even in the  absence of any loops  \cite{nishikawa2006synchronization}.  The simplest such model reads $\dot{\boldsymbol x}_i ={\boldsymbol F}({\boldsymbol x}_i) -\sigma\sum_j L_{ij} {\boldsymbol H} ({\boldsymbol x}_j) $, where  $L=(L_{ij})$ is the Laplacian matrix representing the (possibly directed) network of interactions, and the stability of a synchronous solution ${\boldsymbol x}_i(t)= {\boldsymbol s}(t)$ $\forall i$ is  determined by a master stability function that often has a  bounded stability region \cite{pecora1998master,nishikawa2006maximum}. Accordingly, 
the system is synchronizable for a wider range of 
{\color{black} the} coupling strength  $\sigma$ when the nonidentically-null eigenvalues of the Laplacian matrix $L$ are less scattered; the optimally synchronizable networks are those for which these eigenvalues satisfy $\lambda_2=\dots =\lambda_n$. As shown in References \cite{nishikawa2006synchronization,nishikawa2006maximum}, 
every network that can be spanned from one of its nodes (a necessary condition for stable synchronization to be possible) can also be converted into an optimally synchronizable network by removing edges or reducing edge strengths. In particular, any unweighted oriented tree spanning the entire network is an optimally synchronizable network. Therefore, starting from an arbitrary network that is not synchronizable, one can always turn it into a synchronizable network by pruning the interactions between oscillators \cite{nishikawa2006maximum}.

Because such systems can have a bounded stability region, they 
{\color{black} exhibit}
a number of other counterintuitive effects 
{\color{black} as a result of}
the nontrivial dependence of the eigenvalue spread on the interaction network. For example, it has been shown that otherwise unstable synchronous states can be stabilized by transiently uncoupling the oscillators 
\cite{schroder2015transient, Schroeder2016} (see also {\color{black} \cite{ stilwell2006sufficient,Chen2009onoffcoupling,jeter2015synchronization}).}
In different work, it was shown that an otherwise nonsynchronizable network can become synchronizable not only by removing nodes but also by adding nodes despite the resulting increase in the number of eigenvalues that need to fit inside the stability region \cite{nishikawa2010network}.

\subsection{2.4.~Control by Antagonistic Interventions}

A problem of fundamental interest in network dynamics is the one of preventing the loss of resources by means of interventions that are themselves limited to only further removing resources. As a concrete example, consider a food-web network in which a primary extinction triggers a cascade of secondary extinctions. The question of interest is to design a control intervention that if applied after the first extinction (but before the propagation of the cascade) would prevent the other extinctions. An elementary model to conceptualize the problem is the $n$-species Lotka-Volterra predator-prey model \cite{cohen1990stochastic}:
\begin{equation}
\dot{X_i}=X_i \left(b_i +\sum_j a_{ij}X_j\right),  ~~ i=1,\cdots, n,
\label{eq5}
\end{equation}
where $X_i\ge 0$ is the population of species $i$ and parameter $b_i$ is the growth rate for basal species (those that do not 
  {\color{black} feed}
on others) and the mortality rate for non-basal ones. Upon removal of one species, this system generically has one fixed point at which all other species populations are nonzero when the matrix $A=(a_{ij})$ is invertible  (for simplicity let us ignore the possible presence of a 
nonfixed-point
 attractor in which all other species survive).  

\begin{figure}[t]
\includegraphics[width=4.3in]{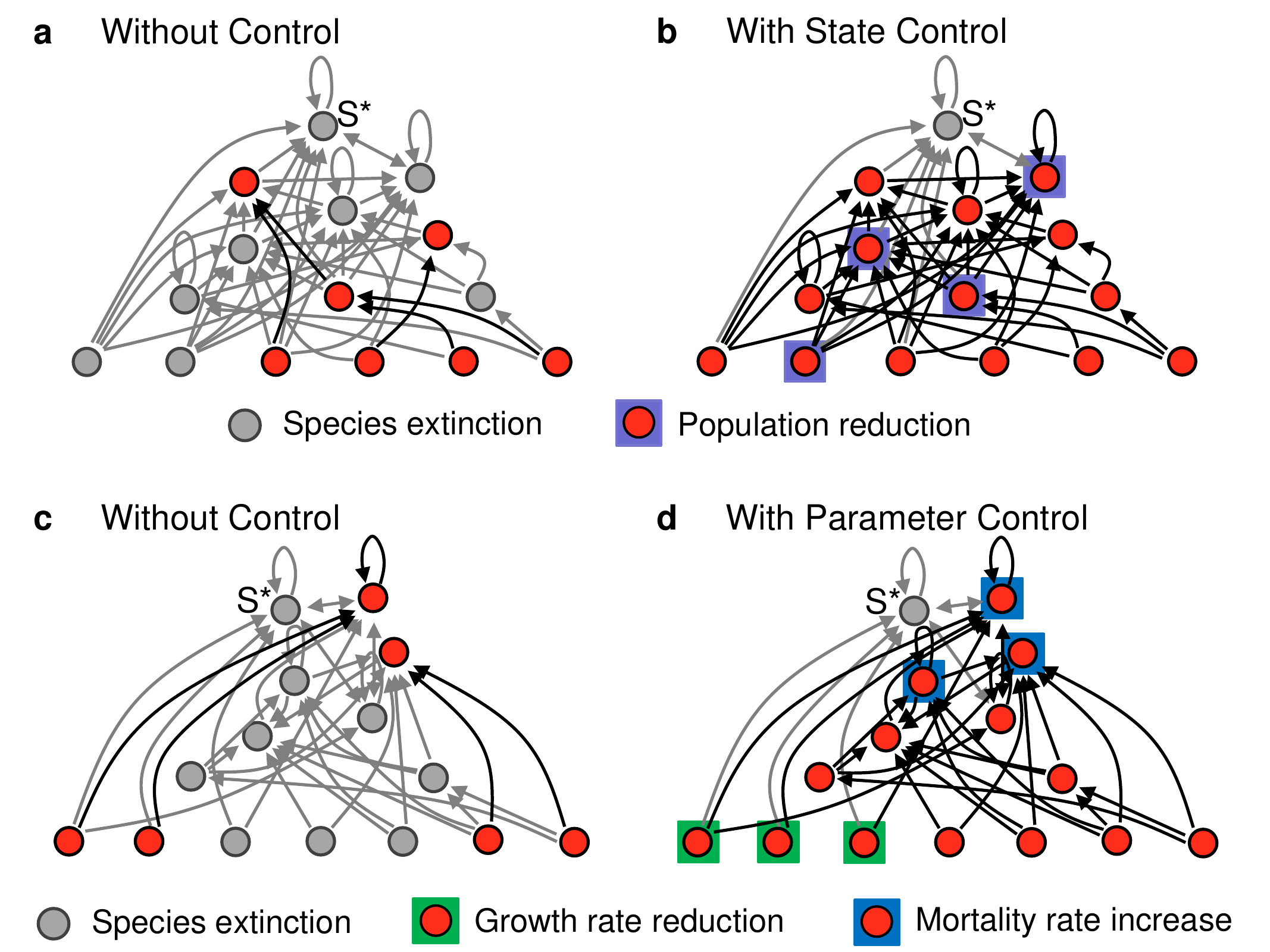} 
\caption{Examples of antagonistic control to mitigate extinction cascades.
State control: 
(a) the sudden extinction of species $S^*$ leads to the subsequent extinction of seven other species; 
(b) all secondary extinctions are prevented by the targeted 
population reduction of four species.
Parameter  control:  
(c) the extinction of species $S^*$  leads to the subsequent extinction of eight other species;
(d) all secondary extinctions are prevented by the targeted reduction of the growth rate of three basal species
and increase of the mortality rate of three non-basal species. 
Figure modified
from  Reference \cite{sahasrabudhe2011rescuing}.
}
\label{fig6}
\end{figure}

The primary extinction will cause subsequent extinctions if 1) this fixed point is unstable and/or outside the positive ($n-1$)-dimensional orthant or 2) the fixed point is stable and in the positive orthant but the initial extinction laid the network state outside its basin of attraction. Within this simplified picture, control to prevent secondary extinctions should 
be geared towards manipulating the position and stability of this fixed point and/or directing the state to its basin of attraction.  Recognizing that realistic interventions over the relevant time scale cannot directly increase species populations, recent research \cite{sahasrabudhe2011rescuing} has 
considered interventions that either temporarily suppress
certain species' populations $\{X_i\}$---to bring the state to the desired basin of attraction---or permanently reduce (increase) their growth (mortality) rates $\{b_i\}$---to manipulate the fixed point and/or basins of attraction.
\textbf{Figure \ref{fig6}} shows for both types of interventions examples in which they would prevent all secondary extinctions for the model in Equation \ref{eq5}.

The scenario just described is ecologically plausible in view of other antagonistic effects that take place in food-web networks, such as the paradox of enrichment \cite{rosenzweig1971paradox}, in which increasing food availability to a prey may eventually lead to the extinction of its predator. More important, the concept generalizes to other network processes, including the control of cascading failures in power-grid networks and the
  {\color{black} reprogramming}
of intracellular networks, where, owing to scenarios similar to the one above,  
the most effective  beneficial interventions often appear to be deleterious  \cite{cornelius2013realistic}.

{\color{black}
\subsection{2.5.~Other More-for-Less Paradoxes in Networks}

The ``inefficiency'' of the Nash equilibrium \cite{dubey1986inefficiency}---that follows from the equilibrium not being globally optimal---has been shown to lead to numerous other 
``more-for-less'' 
paradoxes
 in networks. 
To be specific, we focus on 
variants and implications of the Braess paradox
and note that related  effects can be recognized  across diverse network systems.  

For example, 
in traffic networks exhibiting the Braess paradox,
as originally formulated in Reference
\cite{Braess1969},
the paradox 
has been shown to
actually disappears for sufficiently high traffic demand \cite{nagurney2010negation}. This means that new routes that inadvertently give rise to the paradox may slow traffic when demand is low and not even be used when demand is high. Other work has shown that, in networks with multiple origin and destination nodes, a decrease in demand can in fact lead to an increase in average travel time \cite{fisk1979more,fujishige2017matroids}. An even stronger effect 
has been established in which  an increase in travel time along a route can result in a new equilibrium characterized by the abandonment of a different route connecting the same origin and destination nodes while the original route may continue to be used \cite{steinberg1988prevalence}. In addition, there are also numerous essentially equivalent restatements of the same underlying phenomena,  such as in the conclusion that an increase in the local travel time can lead to a reduction in the global travel time \cite{smith1978road}, or that the overall transport capacity of a network can be reduced upon addition of capacity to individual links \cite{yang1998capacity}.

Such paradoxes have also been considered in numerous concrete settings, both in the context of complex networks \cite{youn2008price,skinner2015price,sole2016congestion} and in specific application domains, including computer networks
\cite{korilis1999avoiding,kameda2000braess},
chemical reaction networks \cite{cornelius2011dispensability,lepore2011computational},  and electric networks \cite{cohen1991paradoxical,skinner2015price,blumsack2007quantitative,nagurney2016physical}.  In electric networks, in particular, it has been shown that for 
graph topologies similar to the one in \textbf{Figure \ref{fig1}b}, the addition of the intermediate (current-carrying) link can create overloads in other links for certain AC circuits \cite{blumsack2007quantitative}  and lead to an increase in voltage drop  for
a fixed current source in  certain two-terminal DC circuits  \cite{cohen1991paradoxical,nagurney2016physical}.

Among physical systems, a major class of applications concerns the study of fluid networks. Using simplified models of the fluid dynamics, it has been shown that increase in the conductivity of individual pipes in a fluid network can lead to increase in power loss, which can be regarded as a fluid analog of resistance. While 
generally not observed for two-terminal networks \cite{calvert1991braess}, 
this behavior has been predicted for single-source multiple-destination delivery networks of both water  \cite{calvert1991braess,keady1995colebrook} and natural gas \cite{andre2010optimization,ayala2013braess}.  
This
behavior can also be characterized as a need 
for
pressure difference 
increases
 to maintain the same outputs 
 following the capacity increase of specific pipes \cite{ayala2013braess}, thus bearing direct analogy with previous results on simple electric circuits \cite{cohen1991paradoxical}. Finally, it is interesting to 
note
that similar transport inefficiency  has recently been observed also in the quantum regime in mesoscopic material networks \cite{pala2012transport,sousa2013braess}, where the  addition of a transport path induces a decrease in the overall conductance.}

\section{3.~INCONGRUOUS COEXISTENCE} 

Consider a network of identical oscillators. It appears intuitive to assume that synchronization into a common state for all oscillators would be facilitated when the interactions between the oscillators are attractive. This assumption is false, however, as it can be shown that in many cases an otherwise unstable synchronous state can be stabilized by turning part of the network interactions repulsive \cite{nishikawa2010network} (this is common, for example, for networks of diffusively-coupled chaotic oscillators with a bounded stability region).
 But why would our intuition suggest the opposite in the first place? 
 One explanation 
 is that we tend to reason in terms of individual interactions---the interaction between an isolated pair of oscillators must indeed  
   {\color{black} be} attractive for them to synchronize.  
Such a local view fails to capture the 
effect that comes from the other interactions in the network, 
suggesting
a situation  
that may be common 
in the study of network dynamics. In this section, we discuss a selection of 
phenomena of significant current interest involving similar (albeit more intricate) 
apparent oxymora.
To keep the discussion focused, we continue to use synchronization dynamics as a main
model process, although we anticipate that many conclusions extend naturally to other 
   {\color{black} forms}
of collective dynamics, including pattern formation, self-organization, herd behavior, 
and consensus processes.

\subsection{3.1.~Converse of Symmetry Breaking}
 
Consider a network of phase oscillators of the form 
\begin{equation}
\dot{\theta}_i=\omega_i +\sigma \sum_j A_{ij} \sin (\theta_j-\theta_i -\alpha),  ~~ i = 1, \cdots, n,
\label{eq6}  
\end{equation}
where $A_{ij}\ge 0$ and $\sigma \cos(\alpha)>0$.  
Each individual oscillator is identified by its natural frequency $\omega_i$, while the other terms represent  interactions between oscillators. What natural frequencies  should the oscillators have in order to facilitate complete synchronization of the form 
{\color{black} ${\theta}_1(t)={\theta}_2(t)=\dots ={\theta}_n(t)$?}
This question can
{\color{black} be}
 answered using a small angle approximation in the vicinity of the synchronous state to obtain 
$\dot{\theta}_i=\omega_i  -\sigma k_i \sin (\alpha) +\sigma \cos (\alpha) \sum_j A_{ij} (\theta_j-\theta_i )$, where $k_i=\sum_j A_{ij}$ is the indegree of node $i$.  The synchronization condition implies $\dot{\theta}_i=\omega_i  -\sigma k_i \sin (\alpha) \equiv \Omega $ $\forall i$ for some constant $\Omega$, which leads to $\dot{\tilde{\theta}}_i =  -\sigma'\sum_j L_{ij}\tilde{\theta}_j$ for $\tilde{\theta}_i = \theta_i -\Omega t$ and $\sigma' = \sigma \cos (\alpha)$.
The synchronous state is stable if and only if all except the identically null eigenvalue of the Laplacian matrix $L$ have positive real 
   {\color{black} parts,}
which is guaranteed to be the case in any network that can be spanned from a node (as  generally assumed in the study of synchronization). On closer examination, 
the actual
condition for synchronization stability is thus that  
 $\omega_i  = \Omega+\sigma k_i \sin (\alpha) $, 
meaning that the natural frequencies of the individual oscillators have to be nonidentical unless the network has identical indegrees for all nodes. This is
intuitive
because,
in order to achieve an identical state, the oscillators need to be nonidentical to compensate for their nonidentical couplings. 
A generalization of this argument can be used to optimize synchronization in complex networks in general \cite{skardal2014optimal}, and analogous results are expected
if a different characteristic of the system
 (e.g., delays, noise level, or coupling strength) is nonuniform across components.  
For a long time, cases involving such compensatory nonuniformities were the only ones
in which differences between the oscillators were found to help minimize differences between their states.

\begin{figure}[t]  
\includegraphics[width=4.2in]{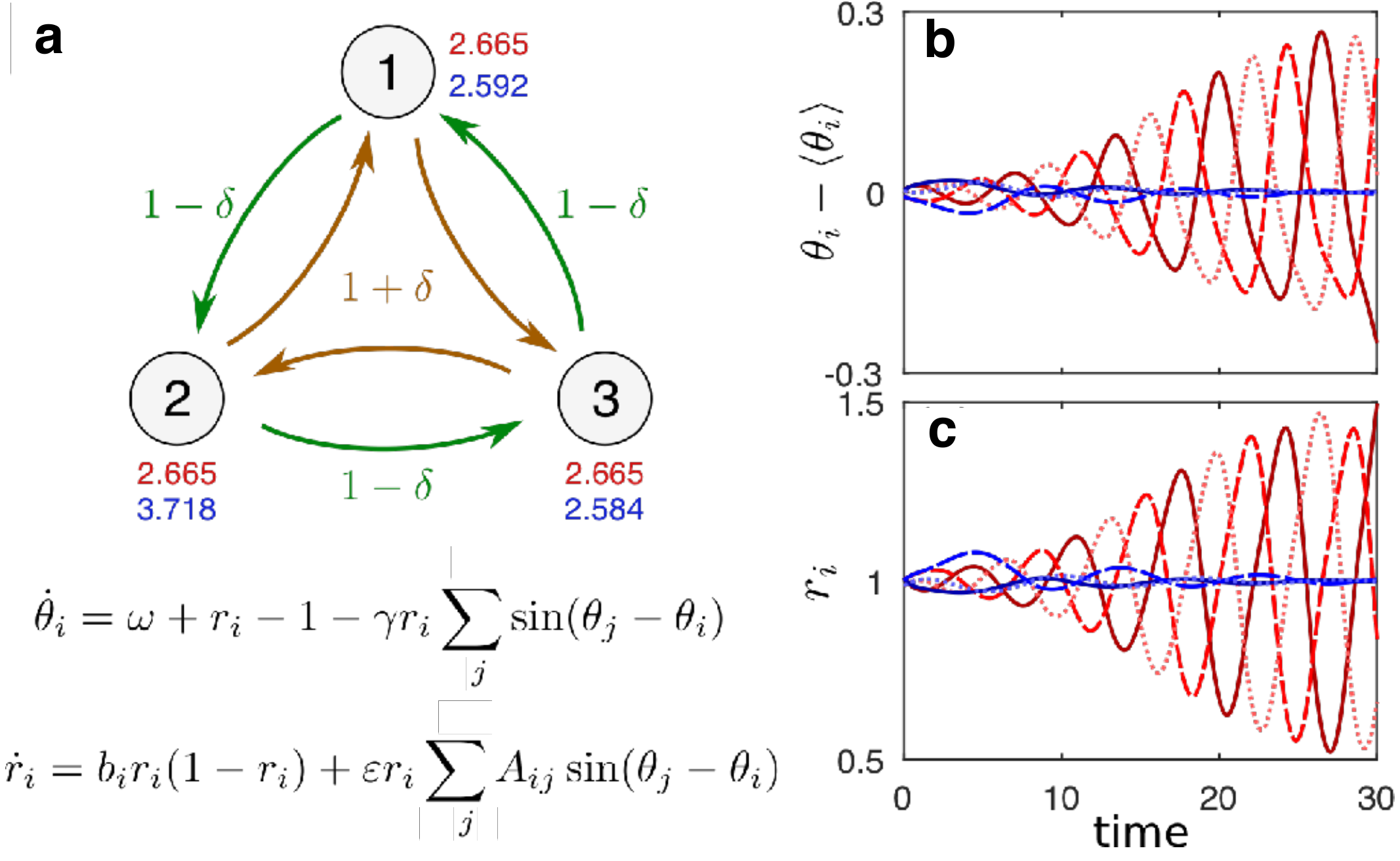}
\caption{Network of identically-coupled oscillators in which the oscillators themselves need to be nonidentical in order to stably synchronize in an identical state.
(a)  Example network of three phase-amplitude oscillators (top), where the edges represent matrix $(A_{ij})$ in the system's equations (bottom).
Marked next to each node are the values of parameter $b_i$ that
optimize synchronization stability with (red) and without (blue) the constraint of 
being identical across the three nodes (for a given homogeneous assignment of all other parameters).
(b, c) Evolution of the oscillators for an initial condition close to the synchronous state, where the red trajectories are for the identical 
parameter assignment,  showing desynchronization, and the blue trajectories are for the  nonidentical  parameter
assignment, which clearly synchronize.
Figure based on Reference \cite{nishikawa2016symmetric}.
}
\label{fig7}
\end{figure}

However, it was recently shown that
stable 
   {\color{black} identical} synchronization can require the oscillators to be nonidentical even when they are identically coupled and indeed equal with respect to all other aspects.  A simple scenario in which this was first demonstrated \cite{nishikawa2016symmetric} was for phase-amplitude oscillators 
characterized by a phase variable $\theta_i$ and an amplitude variable $r_i$ such that the system always has one synchronous solution corresponding 
   {\color{black} to}   
{\color{black} $\theta_1(t)=\theta_2(t)=\dots = \theta_n(t)$ and $ r_1(t)= r_2(t)=\dots =r_n (t)= 1$.}
The question is whether this solution is stable or not. The uncoupled dynamics of the amplitude variable takes the form $\dot r= b_i r(1-r)$, where $ b_i$ is the only parameter allowed to vary from node to node. All other parameters and the network structural properties are identical for all oscillators. As illustrated in \textbf{Figure \ref{fig7}} for a rotationally invariant network of three such oscillators, there are scenarios for which the synchronous state is not stable for any choice of $b_1=b_2=b_3$, but it becomes stable for specific choices of nonidentical $b_i$. This is remarkable because the synchronous state is the state that would reflect the rotational symmetry of the 
  {\color{black} system and, nevertheless, this symmetric state is} 
  stable only when 
   {\color{black} the}
 symmetry of the system is broken by 
making the oscillators nonidentical. 

It is instructive to contrast this effect with the well-studied phenomenon of symmetry breaking. The fact that the symmetric system does not exhibit a stable synchronous (hence symmetric) solution can be seen as an ordinary manifestation of spontaneous symmetry breaking. On the other hand, the fact that the symmetry of the stable solution is preserved when (and only when) the symmetry of the system is broken can be seen as the converse of symmetry breaking. More generally, 
in the same way the former shows that an asymmetric reality may be described by a symmetric theory, the latter shows that a symmetric reality may require an asymmetric theory.

The specific observation that synchronization can be stabilized or enhanced by tuning the oscillators to be nonidentical has potential implications for power-grid networks, whose operation requires frequency synchronization among power generators. As shown in Reference \cite{nishikawa2015stability}, the stability of the synchronous state of interest can be significantly enhanced when an effective parameter that depends on the damping, inertia and droop coefficients of the power generators is set to be suitably different for different generators.  

\subsection{3.2.~Chimera States}

In networks of coupled oscillators, symmetry breaking itself can lead to rather surprising spatiotemporal patterns formed by two or more domains of qualitatively different dynamics, some in which the oscillators are mutually synchronized and 
  {\color{black} others} in which they evolve incoherently. 
First identified by Kuramoto 
\cite{kuramoto2003reduction}
and later termed chimeras \cite{abrams2004chimera}, such states may provide insights into unihemispheric sleep  in some animal species \cite{rattenborg2000behavioral} and fibrillation in the cardiac muscle of ventricles \cite{davidenko1992stationary}.

\begin{figure}[t]  
\includegraphics[width=2.5in]{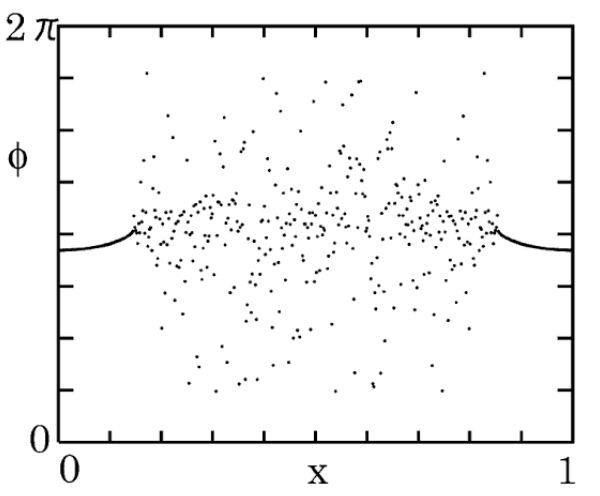}
\caption{
Reproduction of Kuramoto's first published plot of a chimera state, showing the coexistence of a domain of incoherence (scattered points) and a domain of coherence (continuous lines).
This figure was originally published in Reference \cite{kuramoto2003reduction} and 
presents 
the oscillator phase as a function of the position for
 the ring network system described by Equation \ref{eq7}.
}
\label{fig8}
\end{figure}

The first model in which chimera states were systematically described was a ring network of nonlocally coupled phase oscillators \cite{kuramoto2002coexistence},
\begin{equation}
\frac{\partial \phi(x,t)}{\partial t}= \omega - \int G(x-x') \sin \left[ \phi(x,t) - \phi(x',t) + \alpha \right] dx',
\label{eq7}  
\end{equation}
where the kernel $G(x-x')$ is a decreasing function that determines the distance-dependent strength of the coupling. This equation, which can be derived via phase reduction from a nonlocally coupled complex Ginzburg-Landau equation, describes identically-coupled identical oscillators in the limit of
  {\color{black} a}
 large number of oscillators. The complete synchronous state is always a solution and can in fact be stable. But its basin of attraction is not global and the system also exhibits persistent chimera states that are approached for other initial conditions.  Inspection of the phases of the oscillators along the ring, as in the snapshot in \textbf{Figure \ref{fig8}}, shows a clear separation into a domain of incoherence (center) and a domain of coherence (extremes).  Dynamical variants of such  patterns,
 including spiral chimeras in two-dimensional arrays of oscillators, have been known to exist from the very first studies \cite{kuramoto2003rotating}, and experimental demonstrations of chimeras states have been successful on various systems, including networks of coupled electro-optic \cite{hagerstrom2012experimental}, chemical \cite{tinsley2012chimera}, and mechanical \cite{martens2013chimera} oscillators.
 
In part because the original theory to describe chimera states was based on a self-consistent mean-field solution, it was initially believed that such states would not emerge in systems with local or global coupling \cite{abrams2004chimera} and, moreover, that they would be long lived but not permanently stable in finite-size networks \cite{wolfrum2011chimera}. 
{\color{black} Subsequent research demonstrated}
that chimeras can emerge across a surprising range of models and conditions, which include examples of locally and globally coupled systems \cite{panaggio2015chimera}. 
{\color{black} Moreover, while} 
the question of stability remains open for many systems, stable chimera states have 
now been rigorously
identified in finite-size networks of 
chaotic oscillators \cite{cho2017stable}  {\color{black} (see also \cite{suda2015persistent,hart2016experimental, bick2016chaotic}).} 
Yet, previous 
{\color{black} work}
focused exclusively on discrete systems or systems in which the coupling was not strictly local {\color{black} \cite{schmidt2014coexistence},} leaving open the question of whether locally coupled continuous systems could exhibit chimera states (not to be confused with the continuous representation of discrete systems, such as in Equation (\ref{eq7}), whose variables are not spatially continuous, as  illustrated in \textbf{Figure \ref{fig8}}). 
{\color{black} The latter was addressed by recent research, which demonstrated}
 the existence of symmetry-breaking chimera states whose coherent and incoherent phases are analogous to laminar and turbulent phases of a fluid system \cite{nicolaou2017continuous}, thereby revealing an important connection with the classical field of pattern formation.

\subsection{3.3.~Remote Synchronization}

Distant oscillators in a network can synchronize stably even when they are connected exclusively 
through oscillators that are asynchronous. This remote form of cluster synchronization has potential implications for information processing in the brain \cite{nicosia2013remote} and for secure communication \cite{zhang2017incoherence}, and has been recognized in many systems.  For example, in an initially synchronized undirected star network of diffusively coupled chaotic oscillators, an increase in coupling strength can lead to a short wavelength bifurcation that drives the center node off pace but keeps all the other nodes synchronized; this long-known behavior is observed when the stability region is limited, and has been referred to as a drum-head-mode bifurcation~\cite{pecora1998master}  (see also \cite{winful1990synchronized}).

More recently it has been noticed that variants of this behavior  
  {\color{black} extend}
to complex networks in general, largely owing to cluster synchronous states that derive from network symmetry. For example, Reference \cite{nicosia2013remote} considered the system in Equation \ref{eq6} for identical $\omega_i$  
to show that suitable choices of the phase parameter $\alpha$ lead to a frustrated state in which directly connected oscillators do not synchronize whereas certain oscillators that are distant from each other do. The oscillators that synchronize are 
those symmetrically coupled to the network (i.e., in the same symmetry cluster).
This is 
{\color{black} so}
because the equation of motion remains invariant under the action of the symmetry group of the network, meaning that the system admits a synchronous solution among those nodes, which
{\color{black}  in this case is}
stable even when they are not directly connected.

\begin{figure}[t]
\includegraphics[width=5.5in]{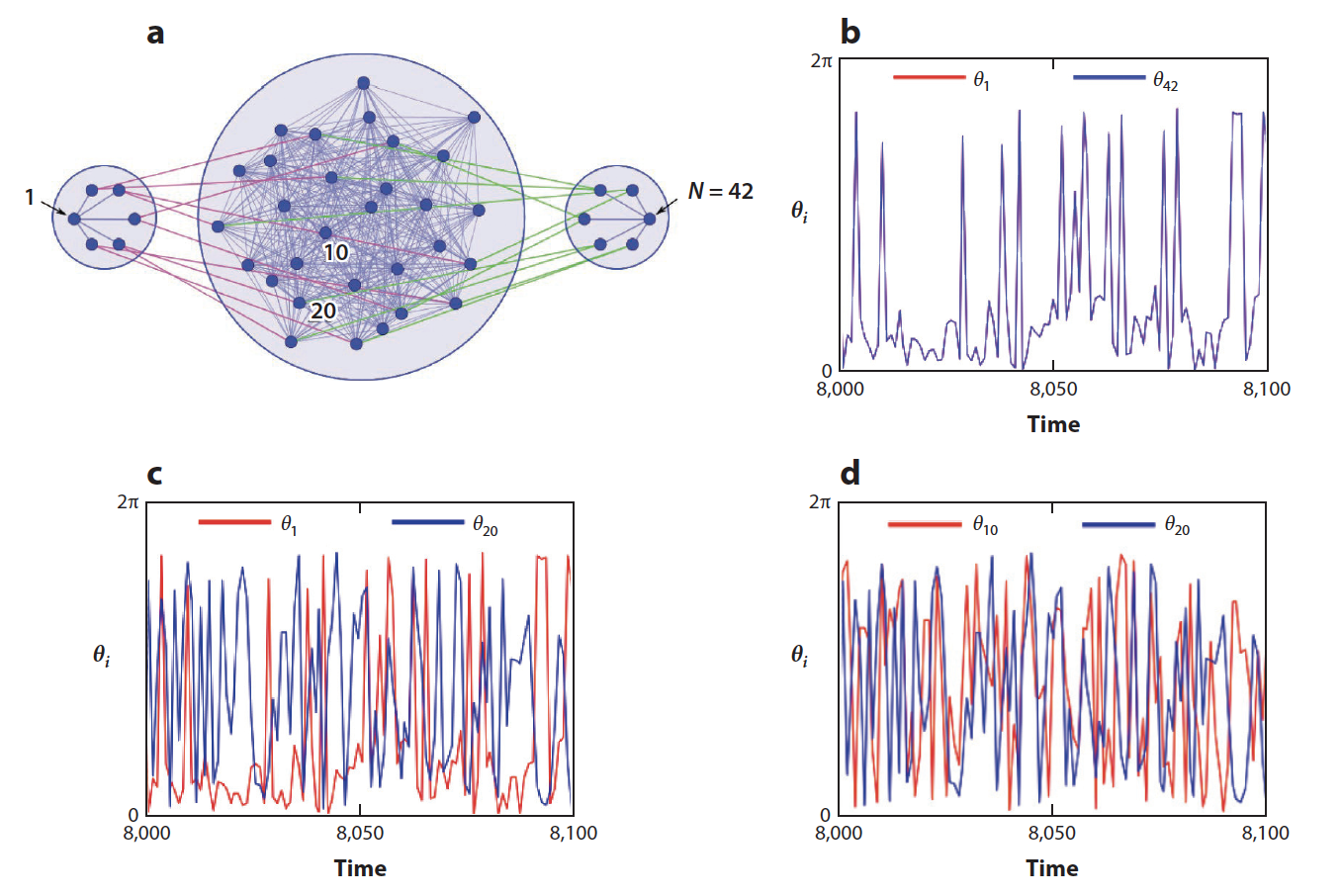}
\caption{Incoherence-mediated remote synchronization. (a) Example of a network equipped with electro-optic phase oscillators. (b) Time evolution of the extreme nodes $1$ and $42$, showing that they are in a state of complete synchronization (the trajectories overlap perfectly). (c, d) Representative plots of the incoherence observed between the extreme nodes and any node in the intermediate group (panel c) as well as between any two nodes in the intermediate group (panel d).
Figure adapted from Reference \cite{zhang2017incoherence}.
}
\label{fig9}
\end{figure}

Several variants of remote synchronization are particularly intriguing. For example, in the study of so-called relay synchronization, it has been shown that two delay-coupled oscillators can synchronize identically (thus without delay) when connected through a third oscillator that lags behind \cite{fischer2006zero}. In a different study, 
two chaotic oscillators have been shown to synchronize stably while connected through an intermediate cluster of oscillators that are incoherent both with the outer nodes and with themselves; termed incoherence-mediated remote synchronization \cite{zhang2017incoherence}, this scenario blends together the properties of remote synchronization with those of chimera states (see \textbf{Figure \ref{fig9}}).  It  has the advantage of being robust to perturbation of the intermediate oscillators and can in principle be useful for encryption key distribution.

A common theme underlying all forms of remote synchronization just mentioned is indeed symmetry in the network. To appreciate how general symmetry-based remote synchronization can be, it is important to notice that even random networks can exhibit a large number of nontrivial symmetry clusters, and that the nodes in such clusters are often not directly connected \cite{macarthur2008symmetry}.  Recently developed techniques \cite{pecora2013symmetries} to study the stability of cluster synchronous states while exploring their relation with symmetries promise to be useful in future studies of this phenomenon.

\subsection{3.4.~Remote Control of Information Routing}

In network systems,
the hallmarks of emergent distributed phenomena are found
not only  in overt manifestations of
collective dynamics  but also  in the associated
information transmission and processing.
These characteristics are common
 across numerous
systems in biology, physics, and engineering, 
 ranging from neural and biochemical circuits to 
self-organized communication networks 
 \cite{Hopfield1982, Cardelli2017, Klinglmayr2012applied, Klinglmayr2012theory}.
In biological systems, in particular,
 information handling 
 is often referred to as  {\it distributed}, but
 how information may be specifically communicated and dynamically routed in such systems is not yet well understood.
Recent 
{\color{black} work~\cite{Kirst2012}}
offers concrete hints on what  distributed
 information routing actually means and what it might 
 condense 
   {\color{black} to,}
 qualitatively and quantitatively. 

A theoretical framework for networks of oscillatory units \cite{Kirst2016}
predicts the patterns of information routing in networks 
 and 
   {\color{black} their}
 dependence on
the interaction network
and other factors.  
The framework is established using model systems of the form
\begin{equation}
\dot{\boldsymbol x}= {\boldsymbol f}({\boldsymbol x}) + {\boldsymbol \xi},
\end{equation}
where ${\boldsymbol f}({\boldsymbol x})$ represents the 
intrinsic time evolution rules and interaction structure of the network and ${\boldsymbol \xi}$ represents an 
external (stochastic) input.
The theory determines
how routing patterns depend on the dynamical reference state
  {\color{black} (taken}
  to be a 
  {\color{black}  periodic or fixed-point}
  solution 
of $\dot{\boldsymbol x}_o= {\boldsymbol f}({\boldsymbol x}_o)$) in the presence of small external inputs. 
Because  ${\boldsymbol x}_o$ and ${\boldsymbol \xi}$ are system-wide variables,
local modifications of individual unit properties,   
network interactions, and  external inputs provide mechanisms to flexibly change information routing throughout the entire network. 

 \begin{figure}[t] 
\includegraphics[width=5.0in]{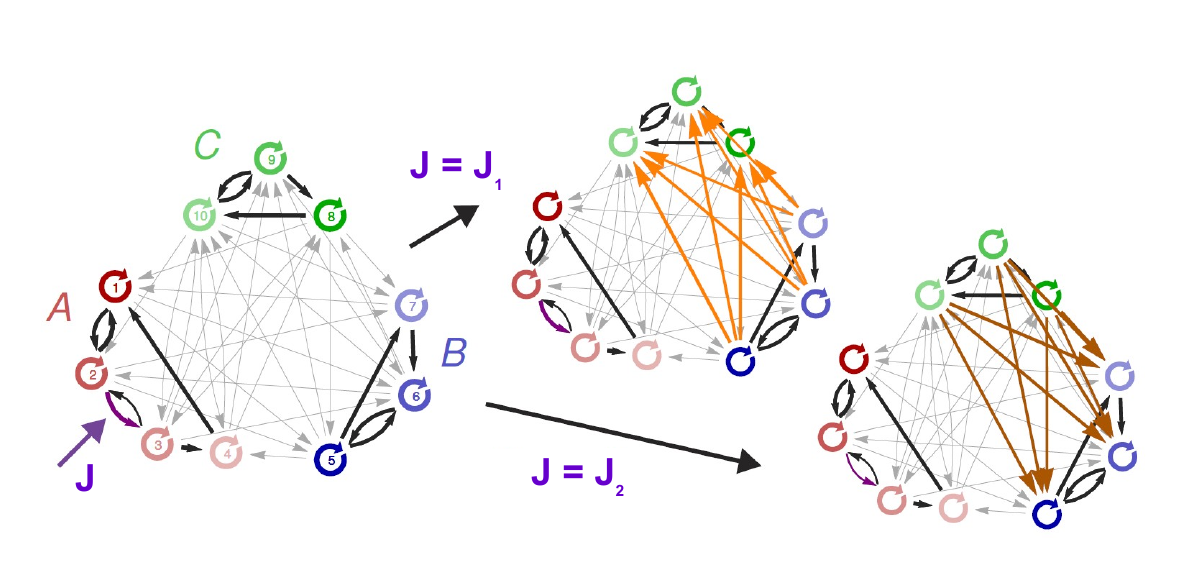}
\caption{
Remote control of information routing in oscillatory networks.
The diagrams illustrate remote control for a hierarchical network, where the uncolored arrows indicate directed 
interactions, which are stronger within modules (black arrows)  than between modules (gray arrows).
Modifying local properties 
(here, a local coupling strength $J$) in one part of the network  (subnetwork $A$) may switch the information routing direction 
between two other parts (subnetworks $B$ and $C$). Specifically, for $J=J_1$, net information is routed from subnetwork $B$ to $C$ 
(orange arrows in top middle panel), whereas for another value of the local coupling strength, $J=J_2$, information is routed from $C$ to $B$ 
(brown arrows in bottom right panel). Figure adapted from Reference \cite{Kirst2016}.
}
\label{fig10}
\end{figure}

As a particularly intriguing property,  we mention that
local modifications in one part of the network
may  remotely influence and
 even reverse the direction of
 information  
 routing
 between two {\it other} parts  (see \textbf{Figure \ref{fig10}}).
 This points to a potentially general mechanism of remote control that
 is possible because information routing is ultimately determined by the network dynamics, which
   {\color{black} are}
 collective {\it and} distributed.

 {\color{black}
 \subsection{3.5.~Other Unconventional Phenomena in Networks}

Limiting ourselves to synchronization dynamics for concreteness, we note that delays, noise, and correlations between node parameters and network structure have been found to lead to other unanticipated phenomena. For example, scenarios have been described in which time delay in node-to-node communication can facilitate rather than inhibit in-phase (zero-lag) synchronization \cite{atay2004delays}. This has been observed, for instance, for pulse-coupled oscillators with inhibitory coupling \cite{ernst1995synchronization} and for relay-coupled oscillators \cite{fischer2006zero}.  For excitatory coupling, on the other hand, networks of pulse-coupled oscillators can be shown to often  exhibit  
attracting periodic orbits
with points that are isolated from their basins of attraction,  
and thus have the peculiar property of being both attractive and unstable \cite{timme2002prevalence}.

Another notable class of behaviors comes from considering coupled oscillators in the presence of noise, where it has been shown that, due to coupling, modular networks can synchronize in response to noise even when the noise applied to different modules is uncorrelated \cite{meng2016independent}. In networks of globally coupled oscillators, different work has demonstrated that independent noise at individual nodes can also stabilize otherwise unstable states of partial synchronization \cite{clusella2017noise}.  Moreover, in systems of directionally coupled non-identical oscillators, it has been shown that the phase diffusion in an oscillator can depend non-monotonically on the noise intensity in a coupled oscillator, and thus become more coherent as noise is further strengthened \cite{amro2015phase}.  

Finally, following the discovery of explosive percolation \cite{achlioptas2009explosive}---characterized by abrupt transitions  \cite{da2010explosive,riordan2011explosive,grassberger2011explosive,nagler2011impact}---analogous transitions  (in fact strictly discontinuous ones \cite{vlasov2015explosive}) have been identified also in synchronization processes, giving rise to so-called explosive synchronization \cite{gomez2011explosive}. Explosive synchronization occurs when, as the coupling strength is increased, an otherwise second-order phase transition  to synchronization becomes first order. This behavior has been demonstrated for both 
phase oscillators \cite{gomez2011explosive} and chaotic oscillators \cite{leyva2012explosive} in 
scale-free networks, where the oscillator frequency is positively correlated with the node degree. Such transitions exhibit hysteresis, in which the 
transformation
from coherence to incoherence occurs at a smaller critical coupling than 
that from incoherence to coherence. The dynamical origin of the hysteretic behavior has been explicitly related 
to
a change in the basin of attraction of the synchronization state \cite{zou2014basin}.
}

\section{4.~OUTLOOK}

What do we learn from these examples  of collective dynamics?
We have illustrated 
{\color{black} various}
types of network
 phenomena, highlighted conditions for their occurrence, and 
identified some common 
{\color{black} mechanisms underlying them.}
Can we {\color{black} expect to} achieve a more unified view 
{\color{black} of}
collective network dynamics in the near future? 

{\color{black}
The question of ``unification'' is indeed a recurrent one in the study of complex systems~\cite{horgan1995complexity}. 
Many argue that the similarities observed in certain phenomena across 
disparate systems are suggestive of common
governing principles.
Others 
further contend 
that an overarching goal of 21st century physics is to construct a unified theory 
of complex systems, which is a pursuit that implicitly assumes that such a unified
account of everything would be both simple {\it and} useful.
The question of whether all observed phenomena can be the result of simple
rules determined by common theories is non-controversial---if we abstract from the fact that we probably do 
not know all fundamental laws of physics, they would all follow from  
a handful of
fundamental 
interactions. But such a description, however simple, is of limited practical
use at the scale relevant to most complex systems phenomena. 
Conversely, computer experiments are useful as broadly applicable approaches to simulate 
the intricate behavior of complex systems but may be no simpler to interpret than the 
empirical data. 
This limitation is all-important
precisely because complex systems tend to
defy  our
ability to understand them. Thus, whether a unified description satisfying the basic 
requirements of simplicity and usefulness can be constructed (even in principle) 
remains
an open question.

On the other hand, networks definitely offer a unified way of thinking about a broad range of 
complex systems. While not all complex systems lend themselves to being described as a network,
the existence of an underlying network of interactions is indeed a defining property of complex systems and 
many complex systems can be usefully represented as a network. 
The network description may not
result in
a theory of everything, but it constitutes a unifying 
principle in and of itself and offers a common framework for the development of broadly
applicable mathematical, computational, and experimental tools, which are 
conducive to 
new discoveries.
Such a description is  sufficiently general to apply to many systems
yet 
sufficiently flexible to account for 
system-specific features as needed. In particular, 
complex systems generally require 
different portrayals
at different scales \cite{anderson1972more}, and networks do offer a versatile representation across scales. 

It appears to be the natural progress of network research that---as more is learned---new  principles 
will be
 discovered,
new tools
will be
developed, and new relations between different systems and phenomena 
will be
established. 
The end product of such a research program, no matter how successful, may not be a unified theory. After all, it may be
argued that there are
principles that govern all, principles that govern some, and principles that govern specific systems. 
This does not reduce the importance of identifying common mechanisms in disparate complex systems.
This article  represents an effort to help 
bring such mechanisms to light and offer a unified view of a broad class of network 
phenomena, even though the systems hosting these phenomena can be very different and with their own idiosyncratic properties in each case.
}

{\color{black}
Looking forward, 
it is  legitimate 
to posit that
further  methodological advances would permit 
development of better understanding
and possibly allow us to} 
predict the limits of, for instance, antagonistic responses or remote actions in networks.
In condensed matter physics,
 {\color{black} for example,}
several innovative  
forms of representing and analyzing collective  
behavior of many-particle systems have 
become standard {\color{black} and now facilitate synergy between subfields.} 
 {\color{black} 
If similar 
overarching 
techniques are developed for
collective dynamics in network systems,} 
they would likely 
be drastically different from
{\color{black} current methods}
in dynamical systems theory,  
where the focus has traditionally been 
{\color{black} on}
 low-dimensional systems.
As illustrated
{\color{black} in various examples of network phenomena considered here,}
the joint presence of  high dimensionality, complex coupling  structure,
and nonlinearity 
{\color{black} leads}
to new phenomena but also pose new challenges.
{\color{black} These examples}
may thus provide some common starting ground to not only  
explore new collective phenomena in their own right but also to develop 
new tools applicable to a broader range of systems and settings. 
{\color{black} Again, developing}
such tools will likely require 
a  shift in theoretical perspective, possibly comparable in significance to the shift required to go from individual particle dynamics  to statistical mechanics.
{\color{black} This shift might, nevertheless, already be underway as a  
co-product 
of the wider adoption of network representations of complex systems.}

{\color{black}
Finally, we note that the phenomena reported here  raise numerous immediate questions for future research.
For instance,
it is instructive to reflect on the more-for-less 
paradoxes
as they relate to the formation and evolution of networks in real systems.
Conceptually, forming a network is often seen as a mode to establish connections, which is a bottom up view that tacitly assumes that the system is built from isolated (or less connected) parts. But a network is also a way to set constraints, which is a top down view that conceptualizes the notion that the system realizes only 
a subset
of all potential interactions.  
The latter
 is relevant in our discussion of network phenomena resulting from the equilibrium state not being the optimal state, since they all are examples in which the  
 state
 realized by the system can be brought closer to the optimum by constraining the structure (or dynamics) of the network.

Still, in real systems,
this alone does not explain why  network resources  whose removal increases performance have not been trimmed over time. 
This question is particularly relevant
in the case of
growing networks, 
such as many biological and infrastructure ones, 
which   
exhibit a net gain rather than loss of
links and nodes as they evolve.
One contributing explanation for this apparent oxymoron is the
pressure imposed by 
the need to perform under
multiple
conditions: while the presence of certain network components may lower performance under the considered condition, they may be needed for improved performance under different conditions. 
 For instance, a living cell activates different parts of its metabolic network depending on the nutrients available in the surrounding medium.
A complementary explanation is that systems often operate under the competing pressures of two or more objectives.
 For example,
 in a power grid the addition of a link to increase power transmission capacity may inadvertently cause desynchronization  \cite{witthaut2012braess}, 
 which illustrates scenarios in which the addition of resources required to improve one function can be strictly deleterious for a different function. The need to understand such scenarios is yet another motivation for future research.
}

A common theme in the network phenomena {\color{black} we}
described
is that, {\color{black} in general, one}
cannot disentangle the network structure from the network dynamics
or attribute the behavior solely to structural properties. This is rooted in the collective and decentralized nature of the dynamics, in which the observed behavior emerges  from interactions.  For example, the relevance of chimera states lies in them being emergent rather than in the mere coexistence of ordered and disordered phases,
which
could be realized by collections of certain bistable oscillators in the absence of any coupling. Much is left for future work, however. 
{\color{black} In particular,} it is important to recognize that,
as much as networks of {\color{black} simple} nodes and links have been powerful in representing a broad range of  complex systems, 
in real systems links and nodes are often complex dynamical systems on their own.

In the broader context of the phenomena illustrated in this review, it is where they run most strongly against our intuition that
we can learn the most and possibly make  the most progress into novel conceptual directions.  We argue for
this perspective of research to make 
advances
into unanticipated network
phenomena currently  unexplained, or still unknown.

\section*{ACKNOWLEDGMENTS}
The authors acknowledge interactions with Daniel Case, Zachary Nicolaou, Thomas Wytock, Christoph Kirst, Jan Nagler, and Takashi Nishikawa, as well as financial support from the National Institutes of Health under grant No.~NIGMS-1R01GM113238 (AEM), the  Army Research Office under award No.~W911NF-15-1-0272 (AEM), the Max Planck Society (MT), 
the Federal Ministry for Education and Research (BMBF) under grant Nos.~03SF0472E and 03SF0472F 
(MT),
{\color{black} and the German Science Foundation (DFG) through a grant
towards the 
Cluster
of Excellence ``Center for Advancing Electronics
Dresden'' (MT).}

\bibliographystyle{ieeetr}
\bibliography{references.bib}

\end{document}